\documentclass[prd, preprint, 12pt]{revtex4-1}

\usepackage{amsmath}
\usepackage{amssymb}
\usepackage{setspace}
\usepackage{graphicx}
\usepackage{natbib}
\usepackage{float}

\begin{document}
 
 % ! TEX spellcheck
 %
%\ \vskip 1.0 in

\begin{center}
 { \large {\bf A new length scale, and  modified Einstein-Cartan-Dirac  equations for a point mass}}

%\smallskip

\vskip 0.1 in

{\large{\bf Tejinder P.  Singh}}

%{\it $^{*}$Indian Institute of Technology Bombay, Powai, Mumbai 400076, India}\\  
{\it Tata Institute of Fundamental Research,}
{\it Homi Bhabha Road, Mumbai 400005, India}\\
%\bigskip
{\tt tpsingh@tifr.res.in}\\

\end{center}

\bigskip
%\bigskip

\centerline{\bf ABSTRACT} 
\noindent We have recently proposed a new action principle for combining Einstein equations and the Dirac equation for a point mass. We used a length scale $L_{CS}$, dubbed the Compton-Schwarzschild length,  to which the Compton wavelength and Schwarzschild radius are small mass and
large mass approximations, respectively. Here we write down the field equations which follow from this action. We argue that the large mass limit yields Einstein equations, provided we assume wave function collapse and localisation for large masses. The small mass limit yields the Dirac equation.  We explain why the Kerr-Newman black hole has the same gyromagnetic ratio as the Dirac electron, both being twice the classical value. The small mass limit also provides compelling reasons for introducing torsion, which is sourced by the spin density of the  Dirac field. There is thus a symmetry between torsion and gravity: torsion couples to quantum objects through Planck's constant $\hbar$ (but not $G$) and is important in the microscopic limit. Whereas gravity couples to classical matter, as usual, through Newton's gravitational constant $G$ (but not $\hbar$), and is important in the macroscopic limit. We construct the Einstein-Cartan-Dirac equations which include the length $L_{CS}$. We find a potentially significant change in the  coupling constant of the torsion driven cubic non-linear self-interaction term in the Dirac-Hehl-Datta equation. We
speculate on the possibility that gravity is not a fundamental interaction, but emerges as a consequence of
wave function collapse, and that the gravitational constant maybe expressible in terms of Planck's constant and the parameters of dynamical collapse models. 

\bigskip
\noindent 
\noindent

%\vskip 1 in

%\centerline{March 31, 2017}
%revised after submitting to GRF

%\bigskip

%\centerline{Essay written for the Gravity Research Foundation 2017 Awards for Essays on Gravitation}
%\bigskip

%\centerline {{\bf Corresponding Author:} Tejinder P. Singh}
\bigskip
%\centerline{\it This essay received an honorable mention in the Gravity Research Foundation 2016 Essay Contest}

%\newpage

\setstretch{1.3}

%\noindent{\it ``It may be that a real synthesis of quantum and relativity theories requires not }
% \noindent{\it just technical developments but radical conceptual renewal''.}
% \qquad\qquad \qquad\qquad\qquad {- J. S. Bell (1986)}
%\bigskip

\section{Introduction}
Compton wavelength and Schwarzschild radius for a point mass $m$  have a peculiar relation to each other, in that their product remains constant at the square of Planck length, as the value of $m$ is changed. Compton wavelength dominates Schwarzschild radius in the quantum regime $m<m_{Pl}$, and vice versa in the classical regime $m>m_{Pl}$.  It seems a reasonable possibility that these two lengths are limiting cases of a unified expression for one length depending on mass, and having a minimum at around
Planck length $L_{Pl}$. The suggestion for such a unified expression appears first in the work of
Carr and collaborators \cite{carr1}, who used it towards investigating the Generalised Uncertainty Principle [GUP] and sub-Planck mass black holes. This has been followed by further work in \cite{Carr2, Carr3, Carr4, Carr5, Carr6}.

In a recent paper \cite{essay2017} we have also argued for such a unified length, which we call the
Compton-Schwarzschild length and which we denote $L_{CS}$. Using this length, we have proposed a common action principle for the general relativistic description of a point mass, and for the Dirac equation. In the present brief note, we write down the field equations following from this action and comment on some of their properties; and in particular the need to include torsion. We consider the minimally coupled Einstein-Cartan-Sciama-Kibble theory (ECSK, and herefater referred to as Einstein-Cartan for brevity). We then propose the modified Einstein-Cartan-Dirac equations motivated by this new length scale $L_{CS}$, and
consider how the idea might be testable through the resulting non-linear Dirac equation.

\section{Einstein-Dirac equations for a point mass}

In \cite{essay2017} we have proposed the following action, in terms of the length $L_{CS}$, to which the
Schwarzschild radius $R_S = 2Gm/c^2$ and half-Compton wavelength $\lambda_C = \hbar/2mc$ are limiting approximations, for $m\gg m_{Pl}$ and $m\ll m_{Pl}$ respectively:
\begin{equation}
\frac{L_{Pl}^2 }{\hbar} S =  \int d^4x \;\sqrt{-g} \left[ R\;  - \frac{1}{2}\; L_{CS}\; \overline{\psi}{\psi} \; + \; L_{CS}^2 \; \overline{\psi}\; i\gamma^{\mu}\partial_\mu\psi 
\right] 
\label{actualaction2}
\end{equation}
We now rewrite this action more accurately, by introducing the desired curved space version of the Dirac kinetic term \cite{Weldon}
\begin{equation}
\frac{L_{Pl}^2 }{\hbar} S =  \int d^4x \;\sqrt{-g} \left[ \frac{1}{8\pi} R\;  - \frac{1}{2}\; L_{CS}\; {\overline\psi}{\psi} \; + \; L_{CS}^2 \; \left\{\frac{i}{2}{\overline\psi}\gamma^{\mu}\; \nabla_\mu\psi  - \frac{i}{2} (\nabla_{\mu}\overline\psi)\gamma^{\mu}\psi\right\}
\; \right] 
\label{actualaction3}
\end{equation}
Variation of this action with respect to the metric, and with respect to  $\overline\psi$, yields the following pair of Einstein-Dirac field equations:
\begin{equation}
R_{\mu\nu} - \frac{1}{2} g_{\mu\nu} R = 8\pi \; L_{CS}^2 \; \chi_{\mu\nu}
\label{gee}
\end{equation} 
\begin{equation}
i\gamma^{\mu}\nabla_\mu \psi = \frac{1}{2 L_{CS}} \psi
\label{gde}
\end{equation}
The symmetric tensor $\chi_{\mu\nu}$ is related to the stress energy-momentum tensor $T_{\mu\nu}$ as
\begin{equation}
\chi_{\mu\nu} = \frac{1}{\hbar c}   \; T_{\mu\nu}
\end{equation}
where $T_{\mu\nu}$, the standard symmetric stress-energy tensor for the Dirac field, is given by Eqn. (3.9) of \cite{Weldon}. 

Let us now consider the large mass and small mass limit of these equations, to ensure that the Einstein equations and the Dirac equation are recovered in the respective limits.
In the large mass limit, consider first the generalised Dirac equation (\ref{gde}), where now $L_{CS}\rightarrow 2Gm/c^2$. So we may write (\ref{gde}) as
\begin{equation}
i\hbar\gamma^{\mu}\nabla_\mu \psi = \left[\frac{\hbar c}{4 G m^2}\right]\; mc\; \psi = \left[\frac{m_{Pl}^2}{4 m^2}\right]\; mc\; \psi 
\label{gde2}
\end{equation}
The factor in brackets on the right hand side results in an enormous suppression to the usual source term
$mc$, when $m\gg m_{Pl}$. For instance, for a mass of 1 gm, there is already a suppression of ten orders of magnitude. Thus it appears reasonable to assume that the amplitude of the state $\psi$ is negligible, except in a very narrow region where the mass $m$ gets localised. This is possible provided we assume that a localisation process, such as collapse of the wave function, is operational, and this has to be described by additional physics not contained in these field equations, and which becomes significant only for large masses. From this suppression factor it is also evident that the Schwarzschild radius is much much larger than the Compton wavelength, which justifies description of the mass $m$ as a black hole. Another way to see this is to note  in the actions (\ref{actualaction2}) and (\ref{actualaction3}) that the kinetic term will be negligible everywhere, upon localisation. The second term also can be assumed to vanish everywhere, except at the point to which the mass $m$ gets  localised. At that point the probability density 
$\overline{\psi}\psi$ given by the second term can be assumed to  the spatial Dirac-delta function $\delta^{3}({\bf x}-{\bf x_0})$, and since  $L_{CS}\rightarrow 2Gm/c^2$, the first and the second term together give the correct Einstein equations for a point mass. This is reaffirmed by recalling that for the Dirac equation in Minkowski space-time, the canonical stress energy-momentum tensor is given by \cite{tong}
\begin{equation}
T^{\mu\nu} = i\hbar c\; \overline{\psi}\gamma^{\mu}\partial^{\nu}\psi
\end{equation}
and hence the total energy is given by
\begin{equation}
E=\int d^{3}x \; T^{00} = \int d^{3}x\; i\hbar c\; \overline{\psi}\gamma^{0}\dot{\psi} = 
\int d^3 x \; \psi^{\dagger} \gamma^{0}(-i\hbar c\; \gamma^i \partial_i + mc^2) \psi
\end{equation}
If the kinetic term can be ignored, and the probability density is replaced by the Dirac-delta function, the total energy is equal to the mass, as desired. A similar result holds in the curved space case, when we consider
the generalised Einstein equations (\ref{gee}). Here, in the large mass limit, the coefficient $L_{CS}^2$ on the right hand side goes to $8\pi\; (2Gm/c^2)^2$. Contrast this with the standard Einstein-Dirac equations, where the coefficient on the right hand side of the Einstein equations is $8\pi G/c^4 \; . \; \hbar c = 8\pi L_{Pl}^2$,
with the $\hbar c$ coming from the $T_{\mu\nu}$. The standard Einstein-Dirac equations {\it do not} go to the point mass black hole limit for large $m$; hence our treatment with an $L_{CS}$ seems to be the more plausible one.

Consider next the small mass limit $m\ll m_{Pl}$ of the Einstein-Dirac equations (\ref{gee},\ref{gde}). In this limit, the coefficient $L_{CS}^2$ on the right hand side of (\ref{gee}) goes to $(\hbar /mc)^2$ and is independent of the gravitational constant $G$. This seems more reasonable than the standard coefficient 
$8\pi G/c^4 \; . \; \hbar c = 8\pi L_{Pl}^2$, for why should $G$ appear at all in the Einstein-Dirac system if 
$m\ll m_{Pl}$, which is essentially the same limit as sending $m_{Pl}\rightarrow \infty$ or equivalently $G\rightarrow 0$. The absence of the coupling constant $G$ from the Einstein-Dirac system suggests to us that in this limit the tensor $\chi_{\mu\nu}$ should not be treated as a source for gravity: in this limit the space-time is Minkowski and the Einstein tensor vanishes. Another way to interpret the situation is that the Schwarzschild radius is now much smaller than the Compton wavelength, and the effect of gravity is very strongly suppressed by the quantum nature of the mass $m$. Since space-time is Minkowski in this limit, the Dirac equation (\ref{gde}) reduces to its flat space-time limit, as desired. 

Nonetheless, we have this curious situation that in this small mass limit the right hand side of Eqn. (\ref{gee}) is non-zero, whereas the left hand side vanishes. One possible way to save the situation is
to introduce a new second rank tensor field on the left side, which, motivated by the reasoning we gave in
\cite{essay2017}, we identify as being due to 
torsion (the anti-symmetric part of the connection). Thus, small masses are a source for torsion, and couple to it through $\hbar$, not through $G$. Gravity is negligible and strongly localised for small masses, whereas torsion is significant. On the other hand, large masses are a source for gravity, and couple to it through $G$, not through $\hbar$. Torsion is negligible and localised for large masses, whereas gravity is
significant. In this manner, we have a symmetry between gravity and torsion. The coupling constant
$L_{CS}^2$ on the right hand side of Einstein equations transits from $(\hbar/2mc)^2$ to $L_{Pl}^2$ to
$(2Gm/c^2)^2$ as the mass changes from $m\ll m_{Pl}$ to $m\sim m_{Pl}$ to $m\gg m_{Pl}$ respectively.
It also appears satisfying that the relative importance of the antisymmetric and the symmetric part of the
connection changes in a harmonious manner as the value of the mass is increased.
The introduction of torsion will modify the action principle (\ref{actualaction3}) - we address this in
the next section. The simplest possibility is that the theory one is looking for is simply the Einstein-Cartan theory \cite{Hehl} with the length scale $L_{CS}$ introduced in it. The action then is a generalisation of the action (\ref{actualaction3}) above, with torsion introduced minimally. We write down these equations in the 
next section, and they seem to have important consequences, because of the introduction of the length scale $L_{CS}$.

{\bf Gyromagnetic ratio}: This is the ratio of the magnetic moment [of a charged body], to its angular momentum. For classical rotating charged bodies this ratio is $\gamma= q/2m$. However, as is well-known, for the Dirac electron, the ratio of its magnetic moment to its spin angular momentum is twice this value:
\begin{equation}
\gamma_e = g. \frac{e}{2m}; \quad \quad g=2
\end{equation}
and we say that the electron has a g-factor of 2. It was first pointed out by Carter \cite{Carter} that the Kerr-Newman black hole \cite{Newman} also has a g-factor of 2. This is somewhat of a surprise, because  black holes are classical objects. However, by expressing mass in terms of the universal length $L_{CS}$ we can easily see why the electron and the Kerr-Newman black hole have the same value for $g$. Since for the electron, we have
$m_e = m_{Pl} L_{Pl} / 2 L_{CS}$, we can write the gyromagnetic ratio as
\begin{equation}
\gamma_e = 2. \frac{e}{2m} = \frac{2e}{m_{Pl}} \; \frac{L_{CS}}{L_{Pl}}
\end{equation}
For a Kerr-Newman black hole, the mass $m_{KN}$ can be expressed in terms of $L_{CS}$ as
$m_{KN}= m_{Pl} L_{CS} / 2 L_{Pl}$, and hence the ratio $\gamma_{KN} = Q/m_{KN}=2. Q/2m_{KN}$
can be written as
\begin{equation}
\gamma_{KN} = \frac{2Q}{m_{Pl}} \frac{L_{Pl}}{L_{CS}}
\end{equation} 
Consider a Kerr-Newman black hole having a charge $Q=e(L_{CS}/L_{Pl})^2$. For such a black hole, the gyromagnetic ratio is
\begin{equation}
\gamma_{KN} =  \frac{2e}{m_{Pl}} \; \frac{L_{CS}}{L_{Pl}}
\end{equation}
We know that the action principle and the field equations above, for a given length $L_{CS}$, describe both a Dirac fermion, as well as a black hole with a dual mass. By comparing the expressions for the gyromagnetic ratio of black hole and electron, we see that a Kerr-Newman black hole with charge $Q$ and mass $m_{KN}$ has the same ratio as a Dirac fermion with dual mass, and dual charge given by
$q = Q (L_{Pl}/L_{CS})^2$. This makes it possible to understand why both have a g-factor of 2: the dynamics of both the systems is described by the same underlying field equations.

\section{The Einstein-Cartan-Dirac equations, and the new length scale}
Let us revisit the action (\ref{actualaction2}) above. We will now generalise it to a curved Riemann-Cartan space-time, possessing both curvature and torsion, and described by the ECSK theory. The curvature scalar now depends on an asymmetric connection, which includes torsion, and the Dirac kinetic term is generalised to a curved background with torsion. We follow the notation and details of the classic review by Hehl et al. \cite{Hehl}. In their notation, as given in their Eqn. (5.11), and keeping in mind the structure of our  actions (\ref{actualaction2}) and (\ref{actualaction3}) above, we write the action for a Dirac field on a Riemann-Cartan background as 
\begin{equation}
\frac{L_{Pl}^2 }{\hbar} S =  \int d^4x \; e \; \left[ \frac{1}{8\pi} R\;  - \frac{1}{2}\; L_{CS}\; {\overline\psi}{\psi} \; - \; i L_{CS}^2 \; \left\{\frac{1}{2}{\overline\psi}\gamma^{\mu}\; \nabla_\mu\psi  - \frac{1}{2} (\nabla_{\mu}\overline\psi)\gamma^{\mu}\psi\right\}
\; \right] 
\label{actualaction4}
\end{equation}
A few comments are in order, as to the structure of this action. It might appear that it is nothing but the standard action, because in the Dirac kinetic term, $L_{CS}$ can be absorbed in the normalisation of the Dirac state. Such a rescaling leaves only the constant $1/L_{CS}$ in front of the mass term as the new input, and this can be trivially redefined as the standard mass term. So what is new in 
(\ref{actualaction4})? There would indeed be nothing new if we ignore the coupling of the Dirac field to gravity [more accurately, to its self-gravity] as represented by the first term in the action integrand, i.e. $R/8\pi$. However, matter-gravity coupling is central to the present discussion, along with the way in which $L_{CS}$ is defined: for large masses it becomes the Schwarzschild radius, and for small masses it becomes Compton wavelength. If we only have a $1/L_{CS}$ in front of the mass term, and nothing in front of the kinetic term, the mass term has the correct coefficient $mc/\hbar$ for small masses, but for large masses it has the wrong coefficient 
$c^2/4Gm$. The mass dependence comes out wrong, and it is this wrong dependence which gets corrected by having an $L_{CS}^2$ in front of the kinetic term, while having an $L_{CS}$ in front of the mass term \cite{essay2017}. The limiting properties of $L_{CS}$ are important.  An equivalent way to see this is to rewrite the action by pulling out an $L_{CS}^2$, as
\begin{equation}
\frac{1}{\hbar}\; \frac{L_{Pl}^2 }{L_{CS}^2} \; S =  \int d^4x \; e \; \left[ \frac{1}{8\pi L_{CS}^2}  R\;  - \frac{1}{2 L_{CS}}\; \; {\overline\psi}{\psi} \; - \; i \; \left\{\frac{1}{2}{\overline\psi}\gamma^{\mu}\; \nabla_\mu\psi  - \frac{1}{2} (\nabla_{\mu}\overline\psi)\gamma^{\mu}\psi\right\}
\; \right] 
\label{actualaction5}
\end{equation}
Now, there is no modification of the kinetic term, but the coupling constant between gravity and the Dirac field has changed from $8\pi L_{Pl}^2$ to $8\pi L_{CS}^2$. This indeed is the essential difference from the standard Einstein-Cartan theory.

Variations of this action with respect to the metric, with respect to the torsion tensor, and with respect to the
Dirac state yields the following system of Einstein-Cartan-Dirac equations with the new length $L_{CS}$ incorporated
in them:
\begin{equation}
G^{ij} = 8\pi \; L_{CS}^2 \;  \frac{1}{\hbar c}\; \Sigma^{ij}
\label{TPS1}
\end{equation}
\begin{equation}
T^{ijk} =8\pi \;  L_{CS}^2 \;  \frac{1}{\hbar c} \; \tau^{ijk}
 \label{TPS2}
 \end{equation}
 \begin{equation}
 i\gamma^{\alpha} \nabla^{\{\}}_{\alpha}\psi + \frac{3}{8} L_{CS}^2 \; (\overline{\psi}\gamma_5 \gamma^{\alpha}\psi) \gamma_5 \gamma_{\alpha}\psi - \frac{1}{2 L_{CS}} \psi =0
 \label{TPS3}
 \end{equation}
 The notation here is the same as in Eqns.  (3.21), (3.22) and (5.14) of \cite{Hehl}. 
 $G^{ij}$ is the asymmetric Einstein tensor; $\Sigma^{ij}$ is the canonical  energy momentum tensor (i.e. the
 asymmetric total energy-momentum tensor which includes also the contribution from the spin angular-momentum tensor $\tau^{ijk}$); and  $T^{ijk}$ is the modified torsion tensor. 
 Our Eqns. (\ref{TPS1}), (\ref{TPS2}) and (\ref{TPS3}) should be compared and contrasted with their Eqns. (3.21), (3.22) and (5.14) respectively. Wherever $L_{Pl}^2$ appears in the original equations, now $L_{CS}^2$ appears  instead: our
 case appears more reasonable, because  now there is no $L_{Pl}$ on atomic or macroscopic scales, and $L_{Pl}$ appears only when $m\sim m_{Pl}$, because then $L_{CS}\sim L_{Pl}$.  We  should also note that a priori we do not know how small masses couple to gravity, and this coupling can only be known by testing a proposal against experiments. Perhaps the most fundamental change has come in the non-linear self-interaction term in the Dirac-Hehl-Datta equation (\ref{TPS3}), where we now have the coupling $L_{CS}^2$ instead of the original $L_{Pl}^2$ in the standard
 Einstein-Cartan theory. The non-linear term is now important at Compton wavelength scales for $m\ll m_{Pl}$, and may
 have significant consequences, as for instance has been suggested in \cite{fabbri}, including perhaps also in the context of ELKO fields \cite{DVA1,DVA2,DVA3}. In a recent work \cite{joy} the authors use torsion in the Dirac equation in an  interesting way to address the hierarchy problem and the divergence of electrostatic and strong force energies for point-like charged fermions. It will be interesting to see how our results in this context compare with those of \cite{joy}. 
 
 In the small mass limit $m\ll m_{Pl}$, where gravity can be neglected and spacetime is flat, and $L_{CS}$ reduces to the half Compton wave-length $\lambda_C$, the Dirac-Hehl-Datta equation  (\ref{TPS3}) becomes
 \begin{equation}
i \gamma^{\alpha} \partial_{\alpha}\psi + \frac{3}{8} \lambda_C^2 \; (\overline{\psi}\gamma_5 \gamma^{\alpha}\psi) \gamma_5 \gamma_{\alpha}\psi - \frac{mc}{\hbar} \; \psi =0
 \label{TPS4}
 \end{equation}
 This equation, which includes a cubic non-linearity in the Dirac equation, should be tested against known
 experimental data, and should also be used to make quantitative predictions which could possibly be verified against new experimental tests. It is well-known that the equation has been used as a model of self-interacting fermions in quantum field theory. However, the so-called `cubic Dirac fermions', which obey such a non-linear equation, have also been studied in condensed matter systems known as `cubically dispersed Dirac semi-metals' \cite{Yang}. The existence of such cubic fermions in `quasi-one-dimensional
transition-metal monochalcogenides' has recently been reported in \cite{Liu} and it would be interesting to
 understand if such materials serve as evidence for the above Eqn. (\ref{TPS4}).

One way to understand the difference between our field equations and the standard Einstein-Cartan equations, for $m\ll m_{Pl}$ is to observe that the change in coupling constant can be represented as
\begin{equation}
\frac{G}{c^4} \; . \; \hbar c = L_{Pl}^2 \rightarrow L_{CS}^2 = \frac{L_{CS}^2} {L_{Pl}^2}\; . L_{Pl}^2 =
     \frac{L_{CS}^2} {L_{Pl}^2}\; \frac{G}{c^4} \; . \; \hbar c 
\end{equation}
 which can be interpreted as
 \begin{equation}
 G\rightarrow G_{new} = \left(\frac{L_{CS}}{L_{Pl}}\right)^2\; G
 \end{equation}
 We might say that gravity has become stronger by the large factor $\left({L_{CS}}/{L_{Pl}}\right)^2$, which is somehow reminiscent of the strong gravity theory proposed by Salam and Strathdee \cite{salam} to address quark confinement. For a proton, this amplification of gravity is by an enormous factor of $10^{38}$. This amplification factor is exactly the same as assumed by Salam and Strathdee, except that they choose the amplification factor in an ad hoc manner, whereas for us it arises naturally in the theory.
 Of course, in reality, $G$ has essentially disappeared from the coupling, and has been replaced by a coupling through $\hbar$. A closer comparison between strong gravity and our small mass limit equations could be worthwhile. 
 
 We recall that in the previous section we were led to include torsion because we expect gravity to vanish
 for $m\ll m_{Pl}$ (the coupling was through $\hbar$, and not $G$). Yet the energy-momentum tensor was
 non-vanishing. Subsequent to the introduction of torsion, we propose the following interpretation. 
 For $m\ll m_{Pl}$ the spin density is more important than mass density: the latter can be neglected
 compared to the former, and the former contributes to torsion. For $m\gg m_{Pl}$, it is the opposite: mass density is more important than spin density, and contributes to gravity. This new interpretation has arisen essentially because we have replaced $L_{Pl}$ in the theory by $L_{CS}$. Torsion and spin density are
 more important in the micro-regime; whereas gravity and mass density are more important in
 the macro-regime.  
 
 Following the discussion in Section V.3 of \cite{Hehl}, we can give a revised estimate as to the relative importance of torsion (spin density) versus gravity (mass density) now that we have the scale $L_{CS}^2$ instead of $L_{Pl}^2$. Associating a spin of $\hbar/2$ with a Dirac fermion of mass $m$, if we consider a continuum fluid of elementary particles with a number density $n$, the mass density is $\rho=mn$, and the spin density is $s=\hbar n/2$. From the field equations (in particular see Eqn. (3.23) of \cite{Hehl}) we can infer that the mass density $\rho$ receives corrections of the order of $\eta\equiv L_{CS}^2\;  s^2 /\hbar c$ from the spin contact interaction. Thus the ratio $\eta / \rho$ is given by
 \begin{equation}
 \frac{\eta}{\rho} = \frac{L_{CS}^2}{ \hbar c }\; \frac{s^2}{\rho} \sim \frac{L_{CS}^2}{ \hbar c } \frac{(\hbar n)^2}
 {mn} \sim n\; L_{CS}^2\; \lambda_C 
 \label{rat}
 \end{equation} 
 Consider first the case $m\ll m_{Pl}$, so that $L_{CS}\rightarrow \lambda_C$. Then, $\eta/\rho \sim  n \lambda_C^3$, implying that spin and hence torsion dominate over mass density and hence over gravity, if there is more than one fermion per Compton volume. Equivalently, torsion becomes more important than gravity when the mass density $\rho$  exceeds $m/\lambda_C^3\sim m^4c^3/\hbar^3$. For neutrons, this is nuclear density
 $\rho \sim 10^{17}$ {kg}/m$^3$. These numbers should be contrasted with those in \cite{Hehl}, and suggest that one should perhaps investigate the significance of torsion for nuclear structure. We note that if $m\rightarrow m_{Pl}$, then spin density dominates mass density only when the mass density approaches Planck density. Thus for sub-Planck masses, the mass density at which torsion becomes more important than gravity decreases as $m$ decreases: for a given mass density, torsion is more important for smaller $m$, as expected.
 
 The results in (\ref{rat}) should be contrasted with those in the standard Einstein-Cartan theory. There, the matter-gravity coupling constant is $L_{Pl}^2$, not $L_{CS}^2$, and the mass term in the Hehl-Datta equation is $\lambda_C$, not $L_{CS}$. As a consequence, the critical radius $r_{crit}$ where torsion becomes comparable to gravity is not $r_{crit}^3 \sim  L_{CS}^2\; \lambda_C$, as we find in (\ref{rat}) above, but rather the so-called Einstein-Cartan radius $r_C$ given by $r_C^3 \sim L_{Pl}^2\; \lambda_C$ (see p. 108 in \cite{Blag}). This is a key difference between the standard theory and our work. One could well ask what $L_{CS}$ is, as regards to its dependence on $m$, for we have only prescribed the asymptotic limits for small and large masses. We believe that the dependence of $L_{CS}$ on $m$, when the mass is not small or large but comparable to Planck mass,  can at present only be determined by experiments. As an illustration. consider the example 
 \begin{equation}
\frac{L_{CS}}{2L_{Pl}} \equiv \frac{1}{2} \left(\frac{2m}{m_{Pl}} + \frac{m_{Pl}}{2m}\right) = \cosh z; \qquad\qquad z\equiv \ln (2m/m_{Pl})
\label{unilength}
\end{equation}
discussed in \cite{essay2017}. For small masses with the mass value approaching Planck mass we can approximate this as
\begin{equation}
L_{CS} \approx \frac{2Gm}{c^2} \left( 1 + \frac {m_{Pl}^2}{4 m^2}\right) 
\end{equation}
This means that if we have two masses $m_1$ and $m_2$ with masses close to Planck mass, their mutual gravitational force is proportional to $Gm_1m_2 (1 + m_{Pl}^2 / 4m_1^2) (1 +  m_{Pl}^2 / 4m_2^2) $. As this example illustrates, the form of $L_{CS}$ could possibly be determined by looking for possible departures from the gravitational force law. A theoretical understanding of the dependence of $L_{CS}$ on $m$ is beyond the present work, and can only come from a microscopic theory of gravitation. The motivation for introducing a new length scale $L_{CS}$ in the Einstein-Cartan theory has been described in some detail in \cite{essay2017}.
 
 Consider next the large mass black hole limit $m\gg m_{Pl}$ where $L_{CS}\rightarrow R_S = 2Gm/c^2$. For a Kerr black hole, the angular momentum $J$ is given by $J=mac$, where $a$ is the Kerr parameter having dimensions of length. Assuming that $V$ is the proper volume inside the horizon radius, we can define a  mass density $\rho = m/V$ and a spin density $s=J/V = mac / V$. Then we have 
 $\eta = L_{CS}^2 s^2 /\hbar c = (2Gm/c^2)^2 (mac)^2 /V^2$ which gives
 \begin{equation}
\frac{\eta}{\rho} \sim \frac{G^2 m^2}{\hbar c^3} a^2 \rho  
\end{equation}  
Assuming $a\sim Gm/c^2$ we can write the condition for spin density to exceed mass density as
\begin{equation}
\rho > \frac {m_{Pl}^2}{m^2} \; \frac{m}{R_s^3} \sim \frac {m_{Pl}^2}{m^2} \; \rho
\end{equation}
which implies that for sub-Planck masses, torsion becomes more important than gravity; also in this case the mass density itself exceeds Planck density. This is expected, and tallies with the result we have in the previous paragraph while approaching Planck mass from below. It also strongly suggests the important role torsion is expected to play in the final stages of gravitational collapse as well as black hole evaporation, possibly avoiding the singularity and replacing it by a bounce, and possibly providing  a resolution of
the black hole information loss paradox.

We also see from the above that the significance of torsion versus gravity  arises differently
in the macro-regime, as compared to the micro-regime. In the former case, Planck mass densities are required for torsion to become important, whereas in the latter case the critical density depends on the 
mass of the elementary particle.
 
 \section{Discussion}
We learn that there is a symmetry between small $m$ and large $m$. The former behaves quantum mechanically, and is a source for torsion. The latter behaves classically, and is a source for gravity. The solution for small mass is dual to the `wave function collapsed' solution for large mass, in the sense that both the solutions have the same value for $L_{CS}$, which is the only free parameter in the theory.

It is evident that gravity emerges only as a result of collapse, when large masses are localised.  There is
no $G$ in the field equations for small $m$. Could it be that the gravitational constant is also not fundamental, but a derived constant, coming from Planck's constant, and possibly from parameters of a wave function collapse model, such as Continuous Spontaneous Localisation (CSL) \cite{rmp}? It is interesting to note that the product $G\hbar$ has mass dimension 0, and has dimensions $[L]^5 \; [T]^{-3}$. We also know that CSL has two free parameters, a rate constant $\lambda_{CSL}$ which determines the rate of wave function collapse, and a length scale $r_C$ which determines the size to which the collapsed wave function is localised in space. Thus one could speculate on a relation such as (in MKS units)
\begin{equation}
G \hbar \sim 10^{-44} \sim r_C^5 \; \lambda_{CSL}^3
\label{Gpara}
\end{equation}
It is evident that the  values for these parameters proposed in the CSL model: $\lambda_{CSL}\sim 10^{-17}$ s$^{-1}$ and $r_C\sim 10^{-7}$ m, do not satisfy this relation. Furthermore, other values of
$\lambda$ and $r_C$ which do satisfy this relation are already ruled out by current experimental bounds \cite{Bassia}. Thus the simplest attempt to relate $G$ to collapse parameters fails. However, it has been pointed out by
Bassi \cite{Bassi2017} that the collapse parameters are phenomenological in nature, and could depend on the environment (strength of the gravitational field present, epoch of evolution of the universe). At this stage we do not necessarily have to conclude that a relation such as (\ref{Gpara}) is impossible to fulfil. The idea that gravity might emerge from collapse has been suggested earlier by Diosi \cite{Diosi}, and also by 
us \cite{si1,si2}, and supports the investigations by other researchers \cite{Paddy, Paddya} that gravity is not a fundamental interaction.

We hope to return to further work on this model, in the future. One would like to examine the non-relativistic weak field limit of the Einstein-Cartan-Dirac field equations, and see if this limit is the same as the Schr\"{o}dinger-Newton equation, or something else. Furthermore, one should try to look for a characteristic solution to these field equations for $m\sim m_{Pl}$. Such a solution ought to have properties `intermediate' between a black hole and a Dirac electron. Some evidence for such `sub-Planck mass' black holes has already been provided in \cite{carr1}. Also, one would like to see if this solution gives rise to a generalised uncertainty principle (GUP) in the spirit of the work of \cite{carr1}. 
Furthermore, in this model, one would like to understand the nature of black hole evaporation in the vicinity of the Planck mass regime \cite{Davidson}, the black hole - elementary particle similarity \cite{Ha}, 
possible evidence for space-time non-commutativity \cite{singha}, quantum gravity induced corrections to the Schr\'{o}dinger equation \cite{singhb}, and the nature of the quantum-classical transition, including in the context of
inflation in the early universe \cite{martin,singhc}. 
It is important also to try and understand how this work relates to dynamical collapse theories of wave function collapse \cite{rmp}. And one would like to explore as to how the torsion field predicted to be significant in the small mass case could be looked for in experiments. The symmetric role of torsion and
gravity is perhaps more general than the context of Dirac fields considered here; a very elegant
illustration in the context of exotic BTZ black holes with torsion has been presented in 
\cite{exotic}. If small masses couple to curvature through torsion rather than through gravity, this could force us to rethink how zero point energy couples to gravity, and could have possible implications for the 
cosmological constant problem and the nature of dark energy.
\bigskip
\bigskip

\noindent{\bf ACKNOWLEDGEMENTS}:  For helpful correspondence, I would like to thank Angelo Bassi, Bernard Carr, Joy Christian, Naresh Dadhich, Aharon Davidson, Fred Diether, Luca Fabbri, Yuan Ha, Friedrich Hehl,  Diego Pavon, and Hendrik Ulbricht. Useful discussions with Sayantani Bera and Srimanta Banerjee are gratefully acknowledged.
 
\bigskip
\bigskip
\bigskip
\newpage

\centerline{\bf REFERENCES}

\bigskip
\setstretch{1.2}

\end{document}